\documentclass{jpsj2}
\usepackage{graphicx}

\title{Unrestricted Hartree-Fock Analysis of Sr$_{3-x}$Ca$_x$Ru$_2$O$_7$}
\author{Shigeru Koikegami\footnote{E-mail: shigeami@secondlab.co.jp}
, Takashi Yanagisawa$^1$, and Soh Koike$^2$}

\inst{Second Lab, LLC, 2-32-1 Umezono, Tsukuba, Ibaraki 305-0045\\
$^1$Electronics and Photonics Research Institute, 
AIST Tsukuba Central 2, Tsukuba, Ibaraki 305-8568\\
$^2$JSPS, Chiyoda, Tokyo 102-8471}

\recdate{\today}

\abst{We investigated the electronic and magnetic structure of Sr$_{3-x}$Ca$_x$Ru$_2$O$_7$ ($0 \leq x \leq 3$) 
on the basis of the double-layered three-dimensional multiband Hubbard model with spin-orbit interaction. 
In our model, lattice distortion is implemented as the modulation of transfer integrals or a crystal field. 
The most stable states are estimated within the unrestricted Hartree-Fock approximation, in which 
the colinear spin configurations with five different spin-quantization axes are adopted as candidates. 
The obtained spin structures for some particular lattice distortions are consistent with the neutron diffraction results 
for Ca$_3$Ru$_2$O$_7$. Also, some magnetic phase transitions can occur due to changes in lattice distortion. 
These results facilitate the comprehensive understanding of the phase diagram of Sr$_{3-x}$Ca$_x$Ru$_2$O$_7$.}

\kword{three-dimensional multiband Hubbard model, double-layered ruthenate, lattice distortion, 
unrestricted Hartree-Fock, spin-orbit interaction}

\begin{document}
\sloppy
\maketitle

%
%
\section{Introduction}
The series of double-layered ruthenates, Sr$_{3-x}$Ca$_x$Ru$_2$O$_7$ ($0 \leq x \leq 3$), 
possesses a variety of phases under a magnetic field ($H$) or pressure ($P$). 
One end-member of the series, Sr$_3$Ru$_2$O$_7$, shows the magnetic field-tuned quantum criticality, 
which is accompanied by the metamagnetic transition around $H\sim 7.85$T.~\cite{Grigera2001} The metamagnetic transition, 
which was initially observed by Cao {\textit et al.}~\cite{Cao1997_1}, was revealed to be a double transition, 
and its second transition in the higher-field is sensitive to the field angle.~\cite{Ohmichi2003} 
Without an applied magnetic field, Sr$_3$Ru$_2$O$_7$ behaves as a Fermi liquid at low temperature.~\cite{SIIkeda2000} 
Angle-resolved photo-emission spectroscopy (ARPES) has been used to observe its well-defined Fermi surfaces,~\cite{Puchkov1998_1} 
and neutron diffraction methods revealed a lack of that long-range magnetic order.~\cite{Huang1998} 
Meanwhile, inelastic neutron scattering has been used to observe two-dimensional ferromagnetic fluctuations 
as incommensurate peaks attributed to Fermi surface nesting.~\cite{Capogna2003} 
These fluctuations induce ferromagnetism when uniaxial pressures along the $c$-axis are applied.~\cite{SIIkeda2001,SIIkeda2004} 
Since neutron diffraction analysis displays the temperature and 
pressure effects on the crystal structure,~\cite{Shaked2000} 
the phase transition induced by uniaxial pressures suggests that the ferromagnetic fluctuations are susceptible to structural changes. 
This situation is similar to the ferromagnetic ground state at the surface of 
Sr$_2$RuO$_4$, as observed by scanning tunneling microscopy (STM).~\cite{Matzdorf2000} 
In Sr$_2$RuO$_4$, the ferromagnetic ground state arises due to a perturbative lattice distortion at the surface, that is, an 
in-plane rotation of the RuO$_6$ octahedron produced by the surface strain.

Another end-member of the series, Ca$_3$Ru$_2$O$_7$, shows two transitions at 
$T_M=48$K and $T_N=56$K.~\cite{Cao1997_2} 
While $T_N$ has been confirmed as an antiferromagnetic ordering temperature,~\cite{Liu1999} 
$T_M$ was believed to be a Mott-like metal-insulator transition temperature though quantum oscillations 
in the $c$-axis resistivity $\rho_c$, for $H \parallel c$ are observed below $T_M$.~\cite{Cao2003_1,Cao2003_2} 
The electrical resistivity and optical conductivity spectra 
of single crystals grown by a floating-zone method proved that the ground state 
of Ca$_3$Ru$_2$O$_7$ is quasi-two-dimensional metallic.~\cite{YYoshida2004,JSLee2004} 
Furthermore, the magnetostriction data for the single crystals demonstrated that 
the first-order transition at $T_M$ can  be attributed to a discontinuous change in the lattice 
constants.~\cite{Ohmichi2004} The quantum oscillations observed in Ca$_3$Ru$_2$O$_7$ 
have also shown that its ground state is metallic with low-carrier density.~\cite{Kikugawa2010} 
The magnetic structure of Ca$_3$Ru$_2$O$_7$ in the ground state was clarified using neutron diffraction analysis: 
the magnetic moments align ferromagnetically within the double layer and antiferromagnetically between 
the double layers.~\cite{YYoshida2005} While these magnetic moments lie along the $b$-axis for $T<T_M$, 
the first-order transition at $T_M$ changes their directions to align with the $a$-axis for 
$T>T_M$.~\cite{Bohnenbuck2008,Bao2008} 
Moreover, the two transitions at $T_N$ and $T_M$ are respectively weakened by the pressures along the $c$-axis 
and those within the $ab$-plane.~\cite{YYoshida2008} These results indicate that 
the magnetic properties of Ca$_3$Ru$_2$O$_7$ are also susceptible to structural changes.

The concentration range $0 < x < 3$ of the Sr$_{3-x}$Ca$_x$Ru$_2$O$_7$ series has been 
investigated in addition to the end members.~\cite{Cao1997_3,SIIkeda1998,Puchkov1998_2,Iwata2008,Qu2008,Qu2009,Peng2010} 
For the intermediate $x$ of this range, the system exhibits a variety of spin structures: 
a cluster spin-glass phase for $0.24 \lesssim x \lesssim 1.2$,~\cite{Iwata2008,Qu2008,Qu2009} 
and a canted antiferromagnetic phase for $1.2 \lesssim x \lesssim 2.0$.~\cite{SIIkeda1998} 
By comparing the Weiss temperatures for $H \parallel ab$ with those for $H \parallel c$, Iwata {\textit et al.}~\cite{Iwata2008} 
elucidate that the magnetic easy axis changes continuously from the $ab$-plane to the $c$-axis with decreasing 
$x$. Peng {\textit et al.}~\cite{Peng2010} confirm that the magnetic easy axis is the $b$-axis for $x=2.4$ and $x=3.0$. 
This result for $x=3.0$ is consistent with the result of a neutron diffraction analysis for 
Ca$_3$Ru$_2$O$_7$.~\cite{YYoshida2005} Moreover, it has been reported that lattice constants vary with 
$x$ in Sr$_{3-x}$Ca$_x$Ru$_2$O$_7$.~\cite{Iwata2008,Peng2010} 

Sr$_{3-x}$Ca$_x$Ru$_2$O$_7$ ($0 \leq x \leq 3$) has also attracted a great deal of theoretical interest. 
The band structures of its end-members, Sr$_3$Ru$_2$O$_7$ and Ca$_3$Ru$_2$O$_7$, have been investigated 
with a local density approximation~\cite{Hase1997} or with the local spin density approximation (LSDA).
~\cite{Singh2001,Singh2006} In particular, the magnetic field-tuned metamagnetic transition of Sr$_3$Ru$_2$O$_7$ 
has been intensively studied as one of the electronic nematic phase transitions on the basis of microscopic theories.
~\cite{Kee2005,Puetter2007,Puetter2010,Raghu2009,WCLee2009,WCLee2010,Fischer2010} The field-induced 
orbital-ordered phase of Ca$_3$Ru$_2$O$_7$ was also investigated on the basis of the spin/orbital model.~\cite{Forte2010} 
However, few theoretical studies have investigated Sr$_{3-x}$Ca$_x$Ru$_2$O$_7$ ($0 < x < 3$),  
while a number of theoretical analyses have been performed for the series of single-layered ruthenates,
~\cite{Nomura2000,Hotta2001,Eremin2002,Kurokawa2002,Mizokawa2004,Okamoto2004,Oguchi2009,Kita2009} 
Ca$_{2-x}$Sr$_x$RuO$_4$ ($0 \leq x \leq 2$), which exhibit the Mott transition at $x \simeq 0.2$.~\cite{Nakatsuji2000} 
The two-dimensional multiband Hubbard model has been utilized by these theoretical analyses of the series of single-layered ruthenates. 
Meanwhile, in order to understand the series of double-layered ruthenates, we need to consider its three-dimensionality. 
The intrinsic importance of the three-dimensionality can be easily found in the experimental results introduced above 
(e.g., the magnetic structure of Ca$_3$Ru$_2$O$_7$). In this paper, we investigate the electronic and magnetic structures of the double-layered ruthenate 
Sr$_{3-x}$Ca$_x$Ru$_2$O$_7$ ($0 \leq x \leq 3$) on the basis of the three-dimensional (3D) multiband Hubbard model. 
Fully considering possible unequivalent sites and the spin-orbit interaction (SOI), 
we determine the ground state of our model for each lattice distortion within the unrestricted Hartree-Fock (UHF) approximation. 
Then, we find that the change in lattice distortion severely affects the electronic and magnetic structures. 
Our results suggest that the many physical phenomena of Sr$_{3-x}$Ca$_x$Ru$_2$O$_7$ ($0 \leq x \leq 3$) have a critical relationship with the change in 
the lattice distortion.

%
%
\section{Formulation}
\begin{figure}
\includegraphics[width=15.6cm]{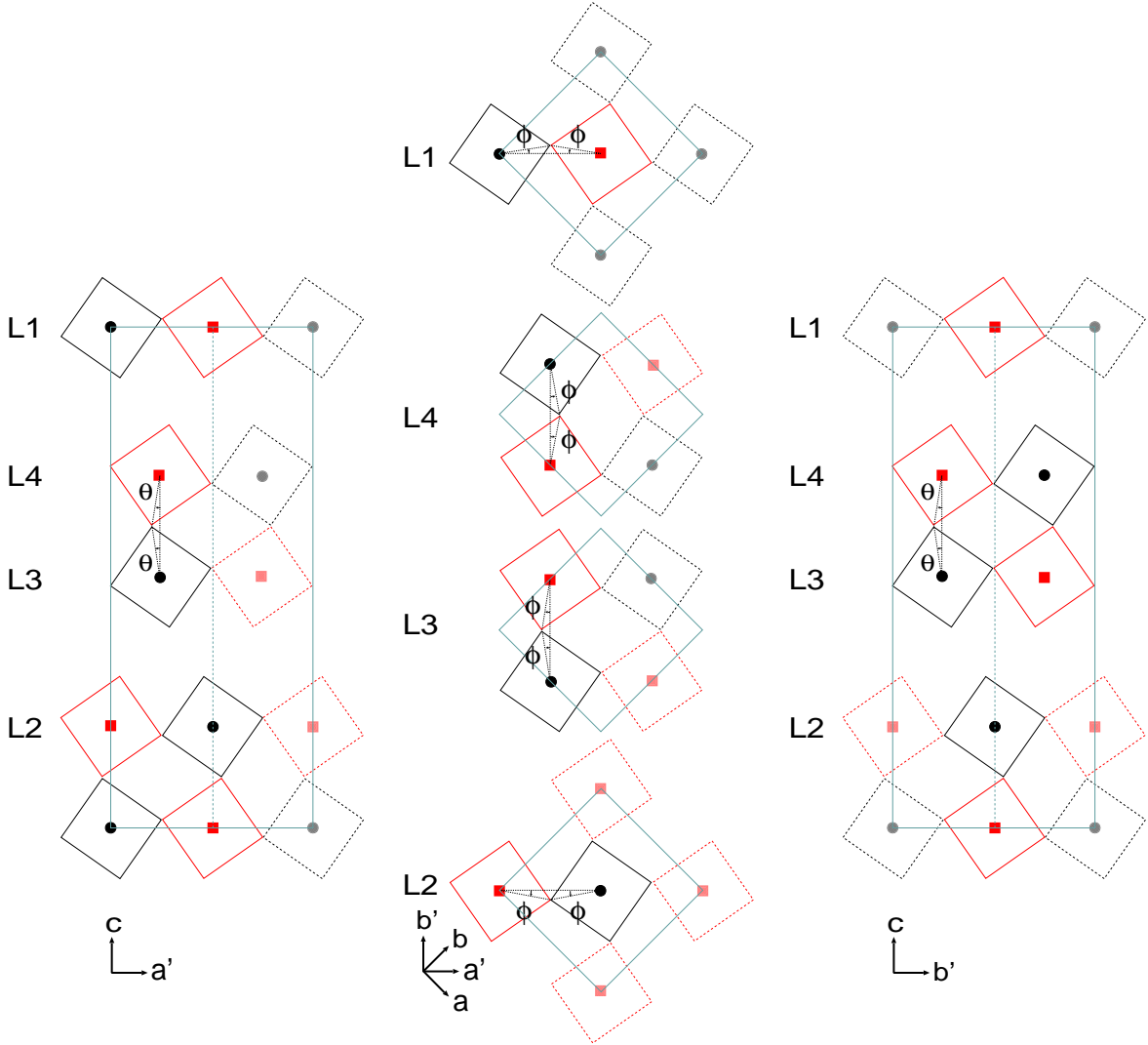}
\caption{\label{figure:1}(Color online) Unit cell of our 3D Hubbard model: 
(left) projection onto the $a^\prime c$-plane, 
(center) projection onto the $a^\prime b^\prime(ab)$-plane, and (right) projection onto the $b^\prime c$-plane. 
The diamonds with solid circles(squares) represent the RuO$_6$ octahedrons on the A(B) sublattice. 
The Ru sites in these octahedrons are indicated by solid circles. L$i$ indicates the 
$i$-th layer ($i=1,2,3,4$). $\phi$ and $\theta$ represent the rotation and tilting angles of the RuO$_6$ octahedron, respectively.}
\end{figure}
Our 3D hubbard model with lattice distortion (Fig.~\ref{figure:1}) consists of A and B sub lattices. 
We consider every three $t_{2g}$ orbitals of the Ru $4d$ electrons located in these sublattices on the $i$-th layer 
($i=1,2,3,4$). 
Thus, our 3D Hubbard model Hamiltonian, $\hat{H}$, is composed as follows:
\begin{eqnarray}
\hat{H} & = & \sum_{i=1}^4\sum_{j=1}^4\sum_{{\mib k}}\sum_{\sigma}\left[\hat{A}^\dagger_{i{\mib k}\sigma}\hat{h}^{AA}_{ij{\mib k}}\hat{A}_{j{\mib k}\sigma}+\hat{A}^\dagger_{i{\mib k}\sigma}\hat{h}^{AB}_{ij{\mib k}}\hat{B}_{j{\mib k}\sigma}+\hat{B}^\dagger_{i{\mib k}\sigma}\hat{h}^{BA}_{ij{\mib k}}\hat{A}_{j{\mib k}\sigma}+\hat{B}^\dagger_{i{\mib k}\sigma}\hat{h}^{BB}_{ij{\mib k}}\hat{B}_{j{\mib k}\sigma}\right] \nonumber \\
& & +\sum_{i=1}^4\sum_{{\mib k}}\sum_{\sigma\sigma^\prime}\left[\hat{A}^\dagger_{i{\mib k}\sigma}\hat{l}^{A}_{\sigma\sigma^\prime}\hat{A}_{i{\mib k}\sigma^\prime}+\hat{B}^\dagger_{i{\mib k}\sigma}\hat{l}^{B}_{\sigma\sigma^\prime}\hat{B}_{i{\mib k}\sigma^\prime}\right] \nonumber \\
& & +\hat{H}^\prime -\mu\sum_{i=1}^4\sum_{{\mib k}}\sum_{\sigma}\left[\hat{A}^\dagger_{i{\mib k}\sigma}\hat{A}_{i{\mib k}\sigma}+\hat{B}^\dagger_{i{\mib k}\sigma}\hat{B}_{i{\mib k}\sigma}\right].
\label{eq:1}
\end{eqnarray}
Here we use the abbreviations 
$\hat{A}^\dagger_{i{\mib k}\sigma}\equiv\left(A^{yz\dagger}_{i{\mib k}\sigma}\,A^{zx\dagger}_{i{\mib k}\sigma}\,A^{xy\dagger}_{i{\mib k}\sigma}\right)$, 
$\hat{A}_{i{\mib k}\sigma}\equiv\,^t\!\left(A^{yz}_{i{\mib k}\sigma}\,A^{zx}_{i{\mib k}\sigma}\,A^{xy}_{i{\mib k}\sigma}\right)$, 
$\hat{B}^\dagger_{i{\mib k}\sigma}\equiv\left(B^{yz\dagger}_{i{\mib k}\sigma}\,B^{zx\dagger}_{i{\mib k}\sigma}\,B^{xy\dagger}_{i{\mib k}\sigma}\right)$, and 
$\hat{B}_{i{\mib k}\sigma}\equiv\,^t\!\left(B^{yz}_{i{\mib k}\sigma}\,B^{zx}_{i{\mib k}\sigma}\,B^{xy}_{i{\mib k}\sigma}\right)$
, where $A^{\varphi}_{i{\mib k}\sigma} (A^{\varphi\dagger}_{i{\mib k}\sigma})$ and $B^{\varphi}_{i{\mib k}\sigma} (B^{\varphi\dagger}_{i{\mib k}\sigma})$ are the annihilation (creation) operators for the electron in the A and B sublattice on the $i$-th layer ($i=1,2,3,4$), as specified by orbital $\varphi=\{yz,zx,xy\}$, momentum ${\mib k}$, and spin $\sigma=\{\uparrow,\downarrow\}$, respectively. $\mu$ is the chemical potential. The nonvanishing $\hat{h}^{AA}_{ij{\mib k}}$, $\hat{h}^{AB}_{ij{\mib k}}$, $\hat{h}^{BA}_{ij{\mib k}}$, and $\hat{h}^{BB}_{ij{\mib k}}$ in eq.~(\ref{eq:1}) are 
\begin{equation}
\hat{h}^{AA}_{ii{\mib k}} = \hat{h}^{BB}_{ii{\mib k}} 
= \left(\begin{array}{ccc}
 0 & \lambda_{\mib k} & 0 \\
\lambda_{\mib k} & 0 & 0 \\
 0 & 0 & \epsilon_{\mib k} 
\end{array}
\right)\,(i=1,2,3,4),
\end{equation}
\begin{equation}
\hat{h}^{AB}_{ii{\mib k}} = \hat{h}^{BA}_{ii{\mib k}} 
= \left(\begin{array}{ccc}
t^{yz}_{ \mib k} & 0 & 0 \\
0 & t^{zx}_{\mib k} & 0 \\
0 & 0 & t^{xy}_{\mib k}
\end{array}
\right)\,(i=1,2,3,4),
\end{equation}
\begin{equation}
\hat{h}^{AB}_{12{\mib k}} = \hat{h}^{BA}_{12{\mib k}} = \left[\hat{h}^{AB}_{21{\mib k}}\right]^* = \left[\hat{h}^{BA}_{21{\mib k}}\right]^* = \hat{h}^{AB}_{34{\mib k}} = \hat{h}^{BA}_{34{\mib k}} = \left[\hat{h}^{AB}_{43{\mib k}}\right]^* = \left[\hat{h}^{BA}_{43{\mib k}}\right]^*
= \left(\begin{array}{ccc}
 c_{\mib k}^z & 0 & 0 \\
 0 & c_{\mib k}^z & 0 \\
 0 & 0 & 0
\end{array}
\right),
\end{equation}
\begin{equation}
\hat{h}^{AA}_{32{\mib k}} = \hat{h}^{BB}_{32{\mib k}} = \left[\hat{h}^{AA}_{23{\mib k}}\right]^* = \left[\hat{h}^{BB}_{23{\mib k}}\right]^* 
= \hat{h}^{AA}_{14{\mib k}} = \hat{h}^{BB}_{14{\mib k}} = \left[\hat{h}^{AA}_{41{\mib k}}\right]^* = \left[\hat{h}^{BB}_{41{\mib k}}\right]^*
= \left(\begin{array}{ccc}
c_{\mib k}^y & 0 & 0 \\
0 & c_{\mib k}^y & 0 \\
0 & 0 & 0 
\end{array}
\right),
\end{equation}
and
\begin{equation}
\hat{h}^{AB}_{32{\mib k}} = \hat{h}^{BA}_{32{\mib k}} = \left[\hat{h}^{AB}_{23{\mib k}}\right]^* = \left[\hat{h}^{BA}_{23{\mib k}}\right]^* 
= \hat{h}^{AB}_{14{\mib k}} = \hat{h}^{BA}_{14{\mib k}} = \left[\hat{h}^{AB}_{41{\mib k}}\right]^* = \left[\hat{h}^{BA}_{41{\mib k}}\right]^*
= \left(\begin{array}{ccc}
c_{\mib k}^x & 0 & 0 \\
0 & c_{\mib k}^x & 0 \\
0 & 0 & 0
\end{array}
\right),
\end{equation}
where we use the abbreviations
\begin{eqnarray}
\label{eq:12}
c^x_{\mib k} & = & -2t_\perp^\prime e^{{\mathrm i}(3k_z/10)} \cos\frac{k_x}{2}, \\
c^y_{\mib k} & = & -2t_\perp^\prime e^{{\mathrm i}(3k_z/10)} \cos\frac{k_y}{2}, \\
c^z_{\mib k} & = & -t_\perp (\cos\phi\cos2\theta)^2 e^{-{\mathrm i}(k_z/5)},
\label{eq:13}
\end{eqnarray}
\begin{eqnarray}
\label{eq:14}
t^{xy}_{\mib k} & = & -2t_1(\cos2\phi\cos\theta)^2\!\left[\cos \frac{k_x+k_y}{2} +\cos \frac{k_x-k_y}{2}\right], \\
\label{eq:17}
t^{yz}_{\mib k} & = & -2t_4\cos \frac{k_x+k_y}{2}-2t_3(\cos\phi \cos2\theta)^2\cos \frac{k_x-k_y}{2}, \\
\label{eq:18}
t^{zx}_{\mib k} & = & -2t_3(\cos\phi\cos2\theta)^2 \cos \frac{k_x+k_y}{2}-2t_4 \cos \frac{k_x-k_y}{2},
\end{eqnarray}
\begin{equation}
\epsilon_{\mib k} = -2t_2\left(\cos k_x+\cos k_y\right)-\Delta, 
\label{eq:16}
\end{equation}
and
\begin{equation}
\lambda_{\mib k} = 2\lambda_0(\cos k_x-\cos k_y).
\label{eq:15}
\end{equation}
In eqs.~(\ref{eq:12})--(\ref{eq:13}) $t_\perp$ and $t_\perp^\prime$ represent inter-layer transfers, and in eqs.~(\ref{eq:14})--(\ref{eq:15}), $t_1$, $t_2$, $t_3$, $t_4$, and $\lambda_0$ represent intra-layer transfers. $\Delta$ in eq.~(\ref{eq:16}) represents the energy level difference between the $d_{xy}$ and $d_{yz}(d_{zx})$ orbitals due to the crystal field. 
The terms $\hat{l}^{A}_{\sigma\sigma^\prime}$ and $\hat{l}^{B}_{\sigma\sigma^\prime}$ in eq.~(\ref{eq:1}), 
arising from the spin-orbit interaction, are determined from formulas that depend on the choice of the spin-quantization axis. Here we only 
consider collinear spin states, with the five different spin-quantization axes as candidates for 
the most stable states. These five are the $c$-, $a^\prime$-, $b^\prime$-, 
$a$-, and $b$-axes. When we consider the state with the spin-quantization axis parallel to the $c$-axis, 
we represent $\hat{l}^{A}_{\sigma\sigma^\prime}$ and $\hat{l}^{B}_{\sigma\sigma^\prime}$ by $\hat{l}^{A(c)}_{\sigma\sigma^\prime}$ and $\hat{l}^{B(c)}_{\sigma\sigma^\prime}$, respectively. These are defined as follows: 
\begin{equation}
\hat{l}^{A(c)}_{\uparrow\uparrow} =  -\hat{l}^{A(c)}_{\downarrow\downarrow} 
= \hat{l}^{B(c)}_{\uparrow\uparrow} = -\hat{l}^{B(c)}_{\downarrow\downarrow}
= \left(\begin{array}{ccc}
0 & \frac{i}{2}\zeta & 0 \\
-\frac{i}{2}\zeta & 0 & 0 \\
0 & 0 & 0 
\end{array}
\right)
\end{equation}
and
\begin{equation}
\hat{l}^{A(c)}_{\uparrow\downarrow} = -\left[\hat{l}^{A(c)}_{\downarrow\uparrow}\right]^* 
= \hat{l}^{B(c)}_{\uparrow\downarrow} = -\left[\hat{l}^{B(c)}_{\downarrow\uparrow}\right]^* 
= \left(\begin{array}{ccc}
0 & 0 & -\frac{1}{2}\zeta \\
0 & 0 & \frac{i}{2}\zeta \\
\frac{1}{2}\zeta & -\frac{i}{2}\zeta & 0
\end{array}
\right).
\end{equation}
Similarly, when we consider the state with the spin-quantization axis parallel to the $a^\prime$-axis, we have  
\begin{equation}
\hat{l}^{A(a^\prime)}_{\uparrow\uparrow} = -\hat{l}^{A(a^\prime)}_{\downarrow\downarrow} 
= \hat{l}^{B(a^\prime)}_{\uparrow\uparrow} = -\hat{l}^{B(a^\prime)}_{\downarrow\downarrow}
= \cos\phi \left(\begin{array}{ccc}
0 & 0 & 0 \\
0 & 0 & \frac{i}{2}\zeta \\
0 & -\frac{i}{2}\zeta & 0
\end{array}
\right)
\label{eq:2}
\end{equation}
and
\begin{equation}
\hat{l}^{A(a^\prime)}_{\uparrow\downarrow} = -\left[\hat{l}^{A(a^\prime)}_{\downarrow\uparrow}\right]^*
= \hat{l}^{B(a^\prime)}_{\uparrow\downarrow} = -\left[\hat{l}^{B(a^\prime)}_{\downarrow\uparrow}\right]^*
= \cos\phi \left(\begin{array}{ccc}
0 & \frac{1}{2}\zeta & -\frac{i}{2}\zeta \\
-\frac{1}{2}\zeta & 0 & 0 \\
\frac{i}{2}\zeta & 0 & 0
\end{array}
\right),
\label{eq:3}
\end{equation} 
and when we consider the state with the spin-quantization axis parallel to the $b^\prime$-axis, we have  
\begin{equation}
\hat{l}^{A(b^\prime)}_{\uparrow\uparrow} = -\hat{l}^{A(b^\prime)}_{\downarrow\downarrow}
= \hat{l}^{B(b^\prime)}_{\uparrow\uparrow} = -\hat{l}^{B(b^\prime)}_{\downarrow\downarrow}
= \cos\phi \left(\begin{array}{ccc}
0 & 0 & -\frac{i}{2}\zeta \\
0 & 0 & 0 \\
\frac{i}{2}\zeta & 0 & 0 
\end{array}
\right)
\label{eq:4}
\end{equation}
and
\begin{equation}
\hat{l}^{A(b^\prime)}_{\uparrow\downarrow} = -\left[\hat{l}^{A(b^\prime)}_{\downarrow\uparrow}\right]^*
= \hat{l}^{B(b^\prime)}_{\uparrow\downarrow} = -\left[\hat{l}^{B(b^\prime)}_{\downarrow\uparrow}\right]^*
= \cos\phi \left(\begin{array}{ccc}
0 & \frac{i}{2}\zeta & 0 \\
-\frac{i}{2}\zeta & 0 & \frac{1}{2}\zeta \\
0 & -\frac{1}{2}\zeta & 0 
\end{array}
\right).
\label{eq:5}
\end{equation}
Furthermore, when we consider the state with the spin-quantization axis parallel to the $a$-axis, we have  
\begin{equation}
\hat{l}^{A(a)}_{\sigma\sigma^\prime} = 
-\frac{1}{\sqrt{2}}(1+\tan\phi)\hat{l}^{A(a^\prime)}_{\sigma\sigma^\prime} 
+\frac{1}{\sqrt{2}}(1-\tan\phi)\hat{l}^{A(b^\prime)}_{\sigma\sigma^\prime} 
\label{eq:6}
\end{equation}
and
\begin{equation}
\hat{l}^{B(a)}_{\sigma\sigma^\prime} = 
-\frac{1}{\sqrt{2}}(1-\tan\phi)\hat{l}^{B(a^\prime)}_{\sigma\sigma^\prime} 
+\frac{1}{\sqrt{2}}(1+\tan\phi)\hat{l}^{B(b^\prime)}_{\sigma\sigma^\prime},
\label{eq:7}
\end{equation}
using eqs.~(\ref{eq:2})- (\ref{eq:5}). 
Here we ignore their $\theta$-dependencies. 
Similarly, when we consider the state with the spin-quantization axis parallel to the $b$-axis, we have  
\begin{equation}
\hat{l}^{A(b)}_{\sigma\sigma^\prime} = 
\frac{1}{\sqrt{2}}(1+\tan\phi)\hat{l}^{A(a^\prime)}_{\sigma\sigma^\prime} 
+\frac{1}{\sqrt{2}}(1-\tan\phi)\hat{l}^{A(b^\prime)}_{\sigma\sigma^\prime} 
\label{eq:8}
\end{equation}
and
\begin{equation}
\hat{l}^{B(b)}_{\sigma\sigma^\prime} = 
\frac{1}{\sqrt{2}}(1-\tan\phi)\hat{l}^{B(a^\prime)}_{\sigma\sigma^\prime} 
+\frac{1}{\sqrt{2}}(1+\tan\phi)\hat{l}^{B(b^\prime)}_{\sigma\sigma^\prime}. 
\label{eq:9}
\end{equation} 
The interacting part $\hat{H}^\prime$ in eq.~(\ref{eq:1}) is represented by 
\begin{eqnarray}
\hat{H}^\prime & = & \frac{U}{2N} \sum_{i=1}^4\sum_{{\mib k} {\mib k}^\prime {\mib q}} 
	\sum_{\varphi}\sum_{\sigma} 
        \left[A_{i{\mib{k+q}} \sigma}^{\varphi\dagger}A_{i{\mib{k^\prime\!-q}} -\!\sigma}^{\varphi\dagger}
	A_{i{\mib{k^\prime}} -\!\sigma}^\varphi A_{i{\mib k} \sigma}^\varphi 
	+B_{i{\mib{k+q}} \sigma}^{\varphi\dagger} B_{i{\mib{k^\prime\!-q}} -\!\sigma}^{\varphi\dagger}
	B_{i{\mib{k^\prime}} -\!\sigma}^\varphi B_{i{\mib k} \sigma}^\varphi \right] \nonumber \\
	 &   & 	\hspace{-0.1em}+\frac{V}{2N} \sum_{i=1}^4\sum_{{\mib k} {\mib k}^\prime {\mib q}} 
	\sum_{\varphi}\sum_{\varphi^\prime\neq\varphi}\sum_{\sigma \sigma^\prime} 
        \left[A_{i{\mib{k+q}} \sigma}^{\varphi\dagger} 
        A_{i{\mib{k^\prime\!-q}} \sigma^\prime}^{\varphi^\prime\dagger}
	A_{i{\mib{k^\prime}} \sigma^\prime}^{\varphi^\prime} A_{i{\mib{k}} \sigma}^\varphi
	+B_{i{\mib{k+q}} \sigma}^{\varphi\dagger}
	B_{i{\mib{k^\prime\!-q}} \sigma^\prime}^{\varphi^\prime\dagger}
	B_{i{\mib{k^\prime}} \sigma^\prime}^{\varphi^\prime} B_{i{\mib{k}} \sigma}^\varphi\right] \nonumber \\
         &   & 	\hspace{-0.1em}+\frac{J}{2N} \sum_{i=1}^4\sum_{{\mib k} {\mib k}^\prime {\mib q}} 
	\sum_{\varphi}\sum_{\varphi^\prime\neq\varphi}\sum_{\sigma \sigma^\prime} 
	\left[A_{i{\mib{k+q}} \sigma}^{\varphi\dagger}
	A_{i{\mib{k^\prime\!-q}}\sigma^\prime}^{\varphi^\prime\dagger}
	A_{i{\mib{k^\prime}} \sigma^\prime}^\varphi A_{i{\mib{k}} \sigma}^{\varphi^\prime} 
        +B_{i{\mib{k+q}} \sigma}^{\varphi\dagger}
	B_{i{\mib{k^\prime\!-q}}\sigma^\prime}^{\varphi^\prime\dagger}
	B_{i{\mib{k^\prime}} \sigma^\prime}^\varphi B_{i{\mib{k}} \sigma}^{\varphi^\prime}\right],\nonumber \\
\label{eq:11}
\end{eqnarray}
where $N$ is the number of ${\mib k}$-space points in the first Brillouin zone (FBZ). 
Here we only consider the on-site interactions, i.e. Coulomb repulsion in the same orbital $U$, 
Coulomb repulsion in different orbitals $V$, and the exchange interaction $J$. 
We adopt the UHF approximation with respect to every two sublattices, four layers, 
three orbitals and two spin states. Thus, we define
\begin{equation}
\frac{1}{N}\left<A_{i{\mib{k}} \sigma}^{\varphi\dagger} 
A_{i{\mib{k^\prime}} \sigma^\prime}^{\varphi^\prime} \right>
\equiv n_{i\varphi \sigma}^A\delta_{{\mib k}{\mib k}^\prime}
\delta_{\varphi \varphi^\prime}
\delta_{\sigma \sigma^\prime}
\end{equation}
and
\begin{equation}
\frac{1}{N}\left<B_{i{\mib{k}} \sigma}^{\varphi\dagger} 
B_{i{\mib{k^\prime}} \sigma^\prime}^{\varphi^\prime} \right>
\equiv n_{i\varphi \sigma}^B\delta_{{\mib k}{\mib k}^\prime}
\delta_{\varphi \varphi^\prime}
\delta_{\sigma \sigma^\prime}.
\end{equation}
Then, we can approximate $\hat{H}^\prime$ as follows:
\begin{eqnarray}
\hat{H}^\prime & \approx & \sum_{i=1}^4\sum_{\varphi} \sum_{\sigma}
	\left[\left\{Un_{i\varphi -\!\sigma}^A
	+\!\sum_{\varphi^\prime \neq \varphi}\left(
	V\sum_{\sigma^\prime}n_{i\varphi^\prime \sigma^\prime}^A
	-J n_{i\varphi^\prime \sigma}^A \right)\right\}\!\left(\sum_{{\mib k}}
	A_{i {\mib k} \sigma}^{\varphi\dagger}
	A_{i {\mib k} \sigma}^\varphi-\frac{N}{2}n_{i\varphi \sigma}^A\right)\right. \nonumber \\
        &   & \hspace{4.4em}+\!\left.\left\{Un_{i\varphi -\!\sigma}^B
	+\!\sum_{\varphi^\prime \neq \varphi}\left(
	V\sum_{\sigma^\prime}n_{i\varphi^\prime \sigma^\prime}^B
	-J n_{i\varphi^\prime \sigma}^B \right)\right\}\!\left(\sum_{{\mib k}}
	B_{i {\mib k} \sigma}^{\varphi\dagger}
	B_{i {\mib k} \sigma}^\varphi-\frac{N}{2}n_{i\varphi \sigma}^B\right)\right]\nonumber \\
\label{eq:10}
\end{eqnarray}
In order to determine the most stable of the five candidates, 
we conduct a self-consistent calculation for each candidate and find the one 
with the lowest electronic energy as estimated by eqs.~(\ref{eq:1}) and (\ref{eq:10}). 
Hence, we can translate the most stable spin-quantization axis as the magnetic easy axis. 
Moreover, for the electronic state that we determine has the most stable spin-quantization axis, 
we can calculate the five types of the magnetic order parameters. Four of these types are antiferromagnetic order parameters, 
referred to as A$_1$-AFM, A$_2$-AFM, C$_1$-AFM, and C$_2$-AFM and expressed as 
\begin{eqnarray}
& & m({\mathrm{A_1}}) = \left|\sum_{i=1}^2(m_i^A+m_i^B)-\sum_{i=3}^4(m_i^A+m_i^B)\right|, \\
& & m({\mathrm{A_2}}) = \left|\sum_{i=2}^3(m_i^A+m_i^B)-\sum_{i=4}^1(m_i^A+m_i^B)\right|, \\
& & m({\mathrm{C_1}}) = \left|\sum_{i=1}^4(-1)^i(m_i^A-m_i^B)\right|, \\
& & m({\mathrm{C_2}}) = \left|\sum_{i=2}^3(m_i^A-m_i^B)-\sum_{i=4}^1(m_i^A-m_i^B)\right|,
\end{eqnarray}
respectively. The fifth type is the ferromagnetic order parameter, expressed as 
\begin{equation}
m({\mathrm{FM}}) = \left|\sum_{i=1}^4(m_i^A+m_i^B)\right|.
\end{equation}
Here, we introduce the magnetic momentum on each site: 
\begin{equation}
m_i^{A(B)} \equiv \frac{1}{2}\sum_\varphi \left[n_{i\varphi \uparrow}^{A(B)}-n_{i\varphi \downarrow}^{A(B)}\right].
\end{equation}
Then we determne the magnetic order with the largest magnitude of these five order parameters for each parameter set.

%
%
\section{Results and Discussion}
In our numerical calculations, we divide FBZ into a $20 \times 20 \times 20$ equally spaced meshes. 
The parameter sets in eqs.~(\ref{eq:12})--(\ref{eq:5}) and (\ref{eq:11}) for these calculations 
are selected as shown in Table~\ref{table:1}. The choice of these parameters is based on preceding theoretical works.~\cite{WCLee2010}
\begin{table*}
\begin{tabular}{ccccccccccccl}
\hline
$t_1$ & $t_2$ & $t_3$ & $t_4$ & $t_\perp$ & $t_\perp^\prime$ & $\lambda_0$ & $\Delta$ & $\zeta$ & $U$ & $V$ & $J$ & \\ 
\hline
$0.40$ & $0.08$ & $0.40$ & $0.04$ & $0.24$ & $0.04$ & $0.04$ & $0.00$ & $0.16$ & $0.80$ & $0.40$ & $0.20$ & Fig.~\ref{figure:2} \\
$0.40$ & $0.08$ & $0.40$ & $0.04$ & $0.24$ & $0.04$ & $0.04$ & $0.16$ & $0.16$ & $0.80$ & $0.40$ & $0.20$ & Fig.~\ref{figure:4} \\
$0.40$ & $0.08$ & $0.40$ & $0.04$ & $0.24$ & $0.04$ & $0.04$ & $0.00$ & $0.16$ & $1.00$ & $0.50$ & $0.25$ & Fig.~\ref{figure:3} \\
$0.40$ & $0.08$ & $0.40$ & $0.04$ & $0.24$ & $0.04$ & $0.04$ & $0.16$ & $0.16$ & $1.00$ & $0.50$ & $0.25$ & Fig.~\ref{figure:5} \\
\hline
\end{tabular}
\caption{\label{table:1}The parameter sets for the calculations. The parameter uint is eV.}
\end{table*}
The ratios $V/U$ and $J/U$ in each set are fixed at $0.5$ and $0.25$, respectively. 
For each set in Table~\ref{table:1}, both the rotation angle $\phi$ and the tilting angle $\theta$ are varied from $0^\circ$ to $20^\circ$ by $5^\circ$. 
All resulting self-consistent fields $n_{i\varphi \sigma}^A$ and $n_{i\varphi \sigma}^B$ 
have four digits of accuracy, and they satisfy 
\begin{equation}
\sum_{i=1}^4 \sum_\varphi \sum_\sigma(n_{i\varphi \sigma}^A+n_{i\varphi \sigma}^B)=32,
\end{equation} 
which means that there are four electrons per Ru site. 

\begin{figure}
\includegraphics[width=15.6cm]{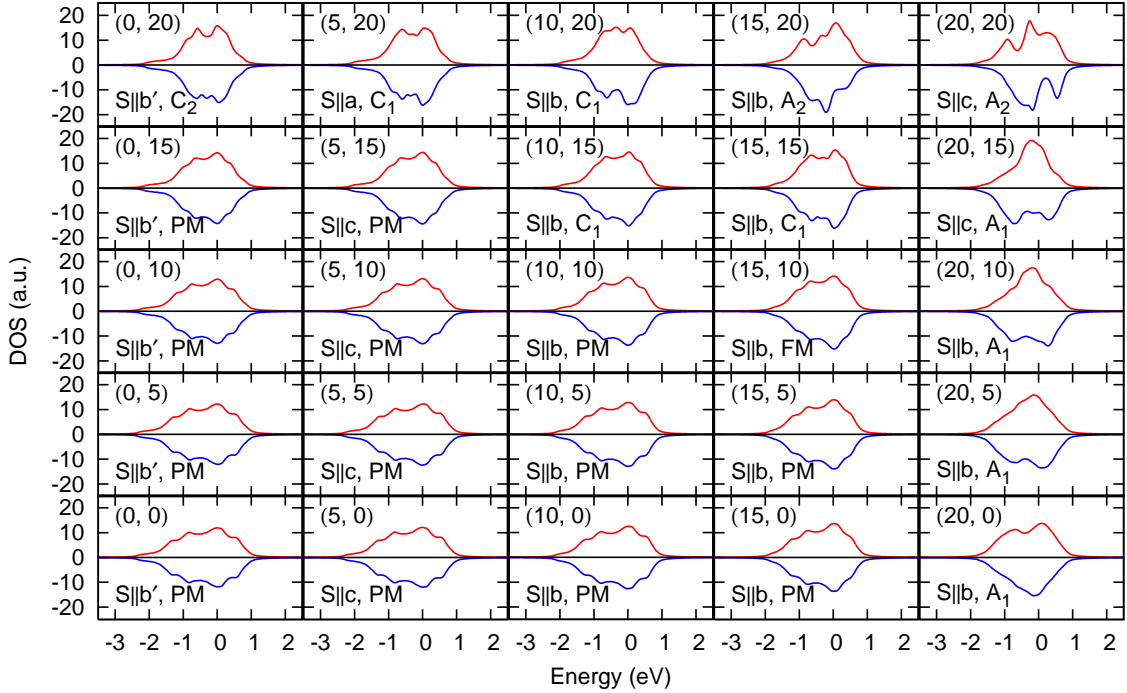}
\caption{\label{figure:2}(Color online) The magnetic easy axis, magnetic phase, and density of states (DOS) determined when $\Delta=0.00\,{\mathrm{eV}}$ and $U=0.8\,{\mathrm{eV}}$. The unit of each $(\phi,\theta)$ is provided in $^\circ$(degrees). Positive DOS is for spin up and negative DOS is for spin down.}
\end{figure}

\begin{figure}
\includegraphics[width=15.6cm]{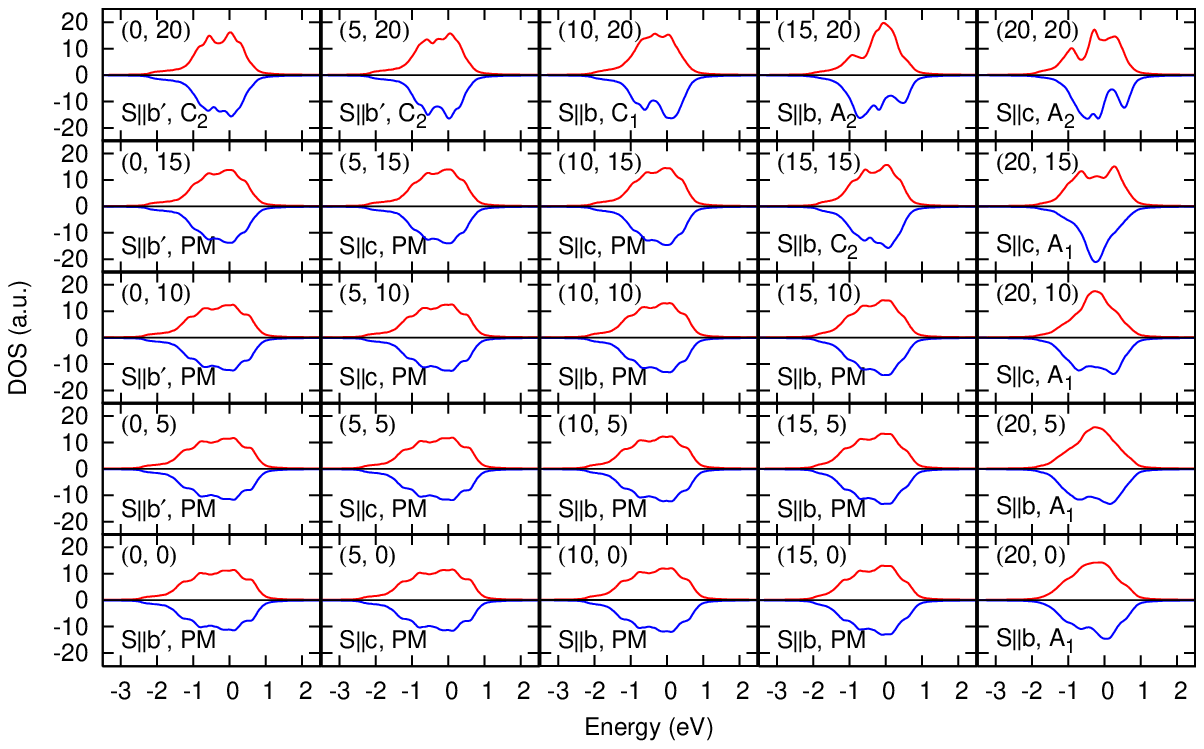}
\caption{\label{figure:4}(Color online) The magnetic easy axis, magnetic phase, and DOS determined when $\Delta=0.16\,{\mathrm{eV}}$ and $U=0.8\,{\mathrm{eV}}$.}
\end{figure}

\begin{figure}
\includegraphics[width=15.6cm]{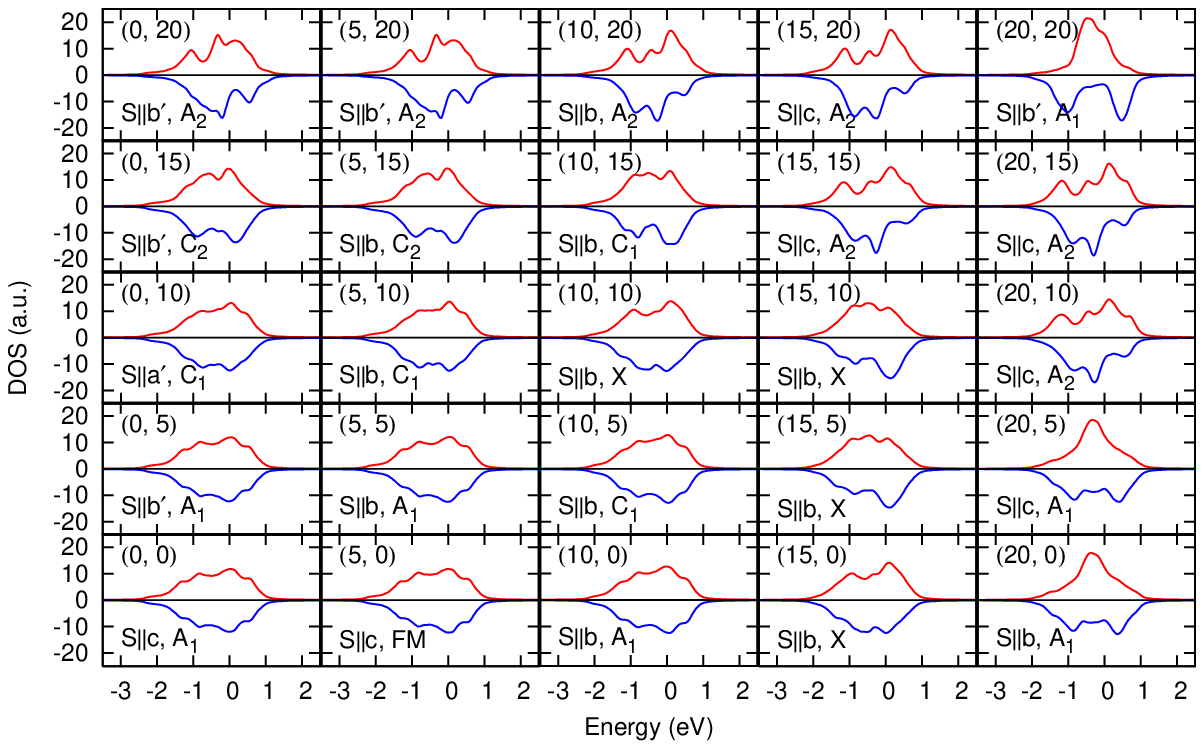}
\caption{\label{figure:3}(Color online) The magnetic easy axis, magnetic phase, and DOS determined when $\Delta=0.00\,{\mathrm{eV}}$ and $U=1.0\,{\mathrm{eV}}$.}
\end{figure}

\begin{figure}
\includegraphics[width=15.6cm]{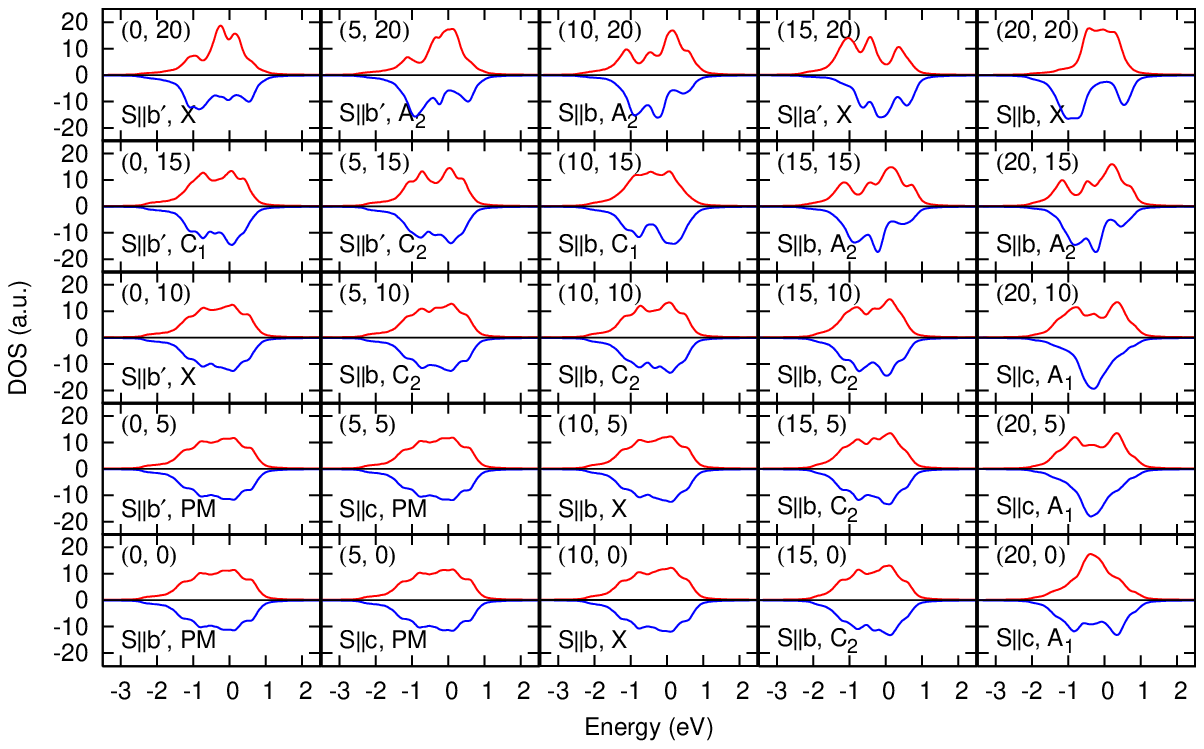}
\caption{\label{figure:5}(Color online) The magnetic easy axis, magnetic phase, and DOS determined when $\Delta=0.16\,{\mathrm{eV}}$ and $U=1.0\,{\mathrm{eV}}$.}
\end{figure}

We summarize our numerical results, as indicated by the last column of Table~\ref{table:1}, in Figs.~\ref{figure:2}-\ref{figure:5}.
The figures show the magnetic easy axis, magnetic phase, and density of states (DOS's) determined for each pair of 
$(\phi,\theta)$.
Here, the magnetic phase is defined as the paramagnetism (PM) when none of the order parameters 
reach a finite value, and the magnetic phase is defined as $\mathrm{X}$ when 
several order parameters simultaneously reach the maximum value of the five. 
We can easily recognize that the various magnetic phases appear and 
that their phase transition is caused by the fractional change of the lattice distortion. 
The magnetic easy axis can vary even in the same magnetic phase. The variations in the magnetic phase and 
easy axis are caused by the transfers and SOI, which both depend on the lattice distortion. 
While the SOI dependence on the lattice distortion plays the primary role in the determination of the magnetic easy axis, 
the transfers dependence is more responsible for determining the magnetic phase. 

The energy level difference between $d_{xy}$ and $d_{yz}(d_{zx})$, i.e. $\Delta$, 
also affects the electronic states 
since it relatively lowers the bands from the $d_{xy}$ orbital as well as enhances the full bandwidth $W$. 
When we compare Fig.~\ref{figure:2} with Fig.~\ref{figure:4}, or 
Fig.~\ref{figure:3} with Fig.~\ref{figure:5}, we find that a positive 
$\Delta$ makes the PM phase stable for a wider range of lattice distortion parameters. 
This can be derived from the decrease in $U/W$ in conjunction with the change in $\Delta$. 
In Ca$_3$Ru$_2$O$_7$, the lattice constants abruptly change at $T_M=48$K, 
where the first-order transition occurs~\cite{Cao2003_1,Ohmichi2004} 
due to Jahn-Teller distortions of the RuO$_6$ octahedra.~\cite{Cao2003_1} 
Thus, we can naturally assign our results for $\Delta > 0$ (Figs.~\ref{figure:4} and \ref{figure:5}) 
to the quasi-two-dimensional metallic state of Ca$_3$Ru$_2$O$_7$ for $T<T_M$.~\cite{YYoshida2004,JSLee2004} 

In our model, all bands can be categorized into two types: 
bands derived from the $d_{xy}$ orbital or bands derived from the $d_{yz}(d_{zx})$ orbital. 
When we respectively define their bandwidths as $W_{xy}$ and $W_{yz,zx}$, 
their dependence on $\phi$ and $\theta$ is like $W_{xy} \propto (\cos 2\phi \cos \theta)^2$ and 
$W_{yz,zx} \propto (\cos \phi \cos 2\theta)^2$, due to eqs.~(\ref{eq:14})-(\ref{eq:18}). 
When either the rotation ($\phi$) or the tilting ($\theta$) increases, both $W_{xy}$ and $W_{yz,zx}$ decrease and $U/W$ increases. 
Thus, the lattice distortion prefers the AFM phase rather than the PM phase due to a large $U/W$, 
as shown in Figs.~\ref{figure:2} and \ref{figure:4}. 
Moreover, each bandwidth dependence creates differences in the lattice distortion effects 
on the electronic state between $\phi$ and $\theta$. The results for 
$(\phi,\theta)=(20,0)$ and $(\phi,\theta)=(0,20)$ in Figs.~\ref{figure:2} and \ref{figure:4} 
provide clear evidence of these differences.

\begin{figure}
\includegraphics[width=11.8cm]{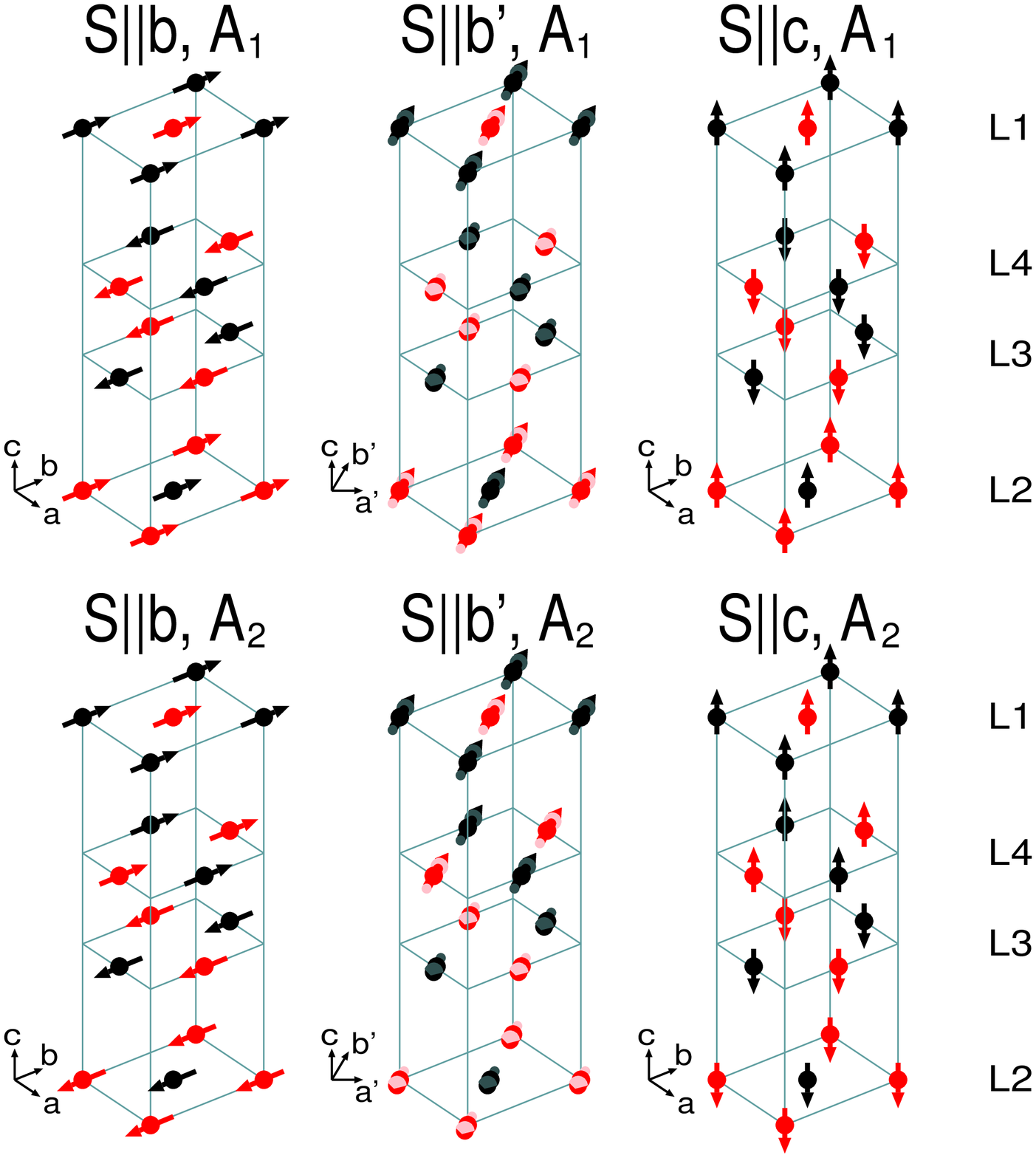}
\caption{\label{figure:6}(Color online) Antiferromagnetic structures obtained in our calculations. The black(red) solid circles represent the Ru site on the A(B) sublattice.}
\end{figure}
\begin{figure}
\includegraphics[width=15.6cm]{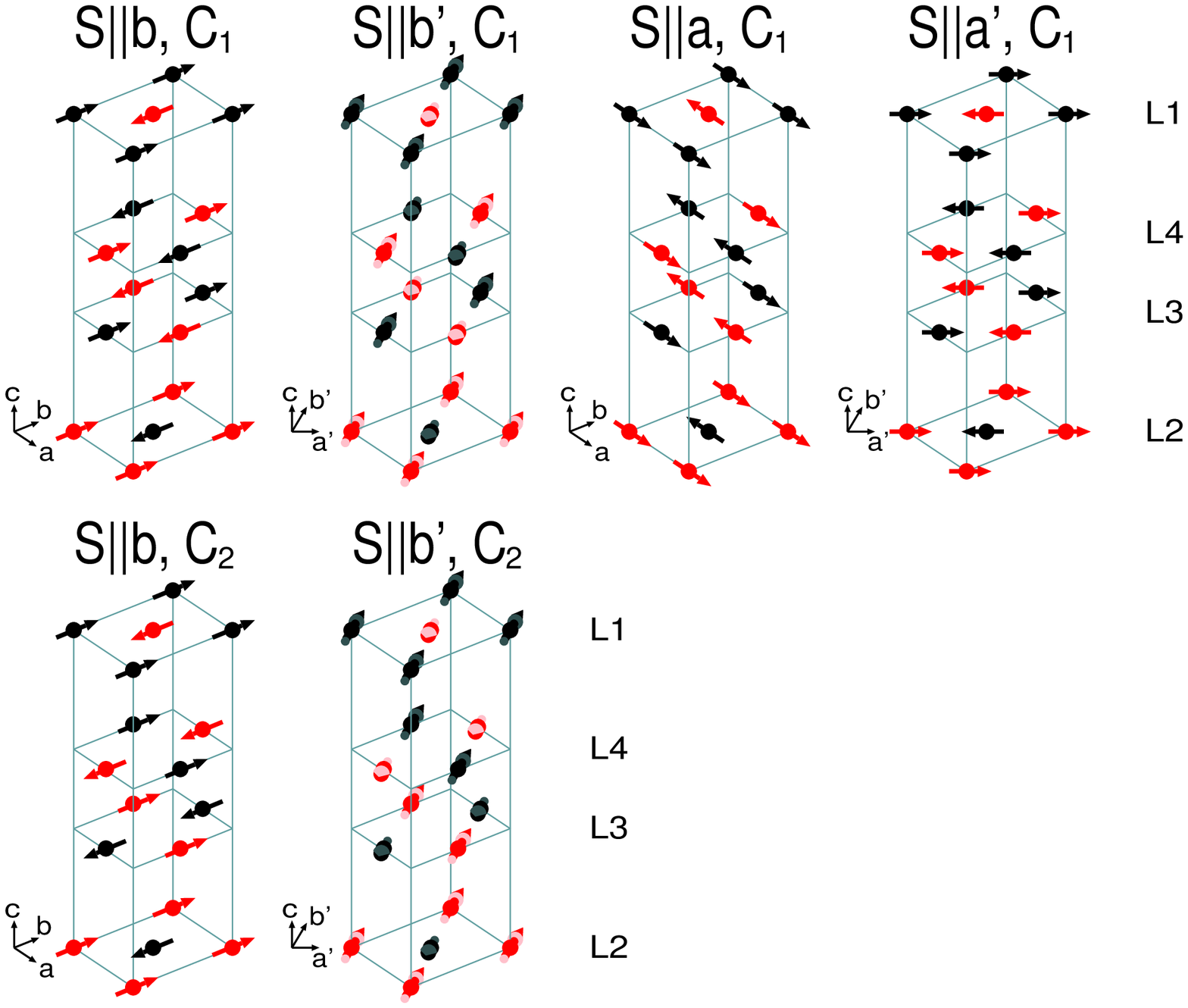}
\caption{\label{figure:7}(Color online) Antiferromagnetic structures obtained in our calculations. The black(red) solid circles represent the Ru site on the A(B) sublattice.}
\end{figure}

Figs.~\ref{figure:6} and \ref{figure:7} show the obtained antiferromagnetic structures. 
In these structures, A$_1$-AFM along the $b$-axis is consistent with the structure of Ca$_3$Ru$_2$O$_7$ 
observed by a neutron diffraction analysis.~\cite{YYoshida2005} The A$_1$-AFM phase appears 
as shown in Figs.~\ref{figure:2}-\ref{figure:4}. This phase has been proved to be 
the most stable state by Singh and Auluck in the LSDA for Ca$_3$Ru$_2$O$_7$, where 
it was noted as AF1.~\cite{Singh2006}
 
The FM phase sometimes appears in the neighborhood of the A$_1$-AFM phase as shown in Figs.~\ref{figure:2} and \ref{figure:3}. 
In the A$_1$-AFM phase, the magnetic moments align ferromagnetically within the double layer, and the ferromagnetic 
correlation within this double layer is stronger than the antiferromagnetic correlation; this 
has been confirmed by an inelastic neutron scattering study for the spin-wave excitation 
in Ca$_3$Ru$_2$O$_7$.~\cite{Ke2011} 
In the intermediate regime where the AFM correlation between the double layer is not fully 
developed, electrons show the FM order as a whole although their magnetic moments are small. 
Thus, FM can be derived from the perturbative change of lattice distortion in our model, 
and this supports the emergence of FM induced by uniaxial pressures along the $c$-axis 
in Sr$_3$Ru$_2$O$_7$.~\cite{SIIkeda2001,SIIkeda2004}

Let us note that the $\mathrm{X}$ magnetic phase appears in Figs.~\ref{figure:3} and \ref{figure:5}. 
Several spin configurations have almost the equivalent energy in this phase, so that 
the energy profile of this phase has a multi-valley structure in its ground state. 
Thus, we can identify the $\mathrm{X}$ magnetic phase 
as a cluster spin-glass phase in Sr$_{3-x}$Ca$_x$Ru$_2$O$_7$ 
for $0.24 \lesssim x \lesssim 1.2$.~\cite{Iwata2008,Qu2008,Qu2009} 
The variation of $x$ in Sr$_{3-x}$Ca$_x$Ru$_2$O$_7$ is accompanied by the change in 
lattice distortion,~\cite{Iwata2008,Peng2010} which must be one reason why the electronic state changes 
with $x$ unless the carrier density remains the same.

%
%
\section{Conclusion}
In this paper, we examined the lattice distortion effects on Sr$_{3-x}$Ca$_x$Ru$_2$O$_7$ 
using the double-layered 3D multiband Hubbard model with SOI 
by the unrestricted Hartree-Fock calculation. 
For some types of lattice distortion, we obtained the A$_1$-AFM phase along the $b$-axis, 
consistent with the neutron scattering result for Ca$_3$Ru$_2$O$_7$. Our results also indicate 
a possible ferromagnetic transition which is susceptible to the change in lattice distortion and the 
existence of a cluster spin-glass phase for the intermediate $x$. The electronic states with these above 
results are all metallic. This suggests that a number of physical phenomena in zero-field can be explained 
without the existence of a metal-insulator transition. In a recent experiment it was reported that the field-induced 
metamagnetic transition drives small lattice distortion in Sr$_3$Ru$_2$O$_7$.~\cite{Stingl2011} 
To elucidate such a relation between lattice distortion and quantum critical phenomena, 
the study of the lattice distortion effects on Sr$_{3-x}$Ca$_x$Ru$_2$O$_7$ in a finite magnetic field is needed in the future.

\section*{Acknowledgments}
The authors are grateful to Drs. Y. Yoshida, S.-I. Ikeda, I. Hase, N. Shirakawa, K. Iwata, and S. Kouno 
for their fruitful discussions. The early stage of our computational study has been achieved 
with the use of Intel Xeon servers at NeRI in AIST.

\end{document}